\documentclass[letterpaper, 10 pt, conference]{ieeeconf}  % Comment this line out
\IEEEoverridecommandlockouts                              % This command is only
\usepackage[utf8]{inputenc}
\usepackage[T1]{fontenc}
\usepackage{graphicx}
\usepackage{amsmath}
\usepackage{amssymb}
\usepackage{url}

\title{\LARGE \bf
Beyond Traditional Surveillance: Harnessing Expert Knowledge for Public Health Forecasting
}

\author{Garrik Hoyt$^{1}$, Eleanor Bergren$^{2}$, Gabrielle String$^{3,4}$, and Thomas McAndrew$^{5}$% <-this % stops a space
\thanks{$^{1}$Department of Computer Science and Engineering, PC Rossin College of Engineering and Applied Sciences, Lehigh University, Bethlehem, PA, USA}%
\thanks{$^{2}$Council of State and Territorial Epidemiologists, Atlanta, Georgia, USA}%
\thanks{$^{3}$Department of Population Health, College of Health, Lehigh University, Bethlehem, PA, USA}%
\thanks{$^{4}$Department of Civil and Environmental Engineering P.C. Rossin College of Engineering and Applied Sciences, Lehigh University, Bethlehem, PA, USA}%
\thanks{$^{5}$Department of Biostatistics and Health Data Science, College of Health, Lehigh University, Bethlehem, PA, USA}%
}

\begin{document}

\maketitle
\thispagestyle{empty}
\pagestyle{empty}

%%%%%%%%%%%%%%%%%%%%%%%%%%%%%%%%%%%%%%%%%%%%%%%%%%%%%%%%%%%%%%%%%%%%%%%%%%%%%%%%
\begin{abstract}

Downsizing the US public health workforce throughout 2025 amplifies potential risks during public health crises. Expert judgment from public health officials represents a vital information source, distinct from traditional surveillance infrastructure, that should be valued---not discarded. Understanding how expert knowledge functions under constraints is essential for understanding the potential impact of reduced capacity.

To explore expert forecasting capabilities, 114 public health officials at the 2024 CSTE workshop generated 103 predictions plus 102 rationales of peak hospitalizations and 114 predictions of influenza H3 versus H1 dominance in Pennsylvania for the 2024/25 season. We compared expert predictions to computational models and used rationales to analyze reasoning patterns using Latent Dirichlet Allocation. Experts better predicted H3 dominance and assigned lower probability to implausible scenarios than models. Expert rationales drew on historical patterns, pathogen interactions, vaccine data, and cumulative experience.

Expert public health knowledge constitutes a critical data source that should be valued equally with traditional datasets. We recommend developing a national toolkit to systematically collect and analyze expert predictions and rationales, treating human judgment as quantifiable data alongside surveillance systems to enhance crisis response capabilities.

\end{abstract}

%%%%%%%%%%%%%%%%%%%%%%%%%%%%%%%%%%%%%%%%%%%%%%%%%%%%%%%%%%%%%%%%%%%%%%%%%%%%%%%%
\section{INTRODUCTION}

Effective public health decision-making requires rigorous training in population-level epidemiology and biostatistics, adherence to evidence-based decision making rather than anecdotal reasoning, and collaborative leadership among networks of health institutions and officials~\cite{c1}. Starting February 14, 2025, the United States Government has---via the canceling of COVID-era grants, policy shifts, and rebudgeting---reduced staff positions related to public health services~\cite{c2,c3}. The US health secretary terminated 10,000 positions within the Department of Health and Human Services. All seventeen experts who were members of the Advisory Council On Immunization Practices were removed from their position~\cite{c4}. The Director of the National Institute of Allergy and Infectious Diseases was put on leave~\cite{c5}. A consistent decline in practitioners, scientists, and epidemiologists associated with public health has been suggested to have a potentially immense impact on US public health~\cite{c6,c7}. The reductions significantly diminished the public health workforce of practitioners, scientists, and epidemiologists, redirecting resources from long-term priorities, such as data collection, to immediate needs~\cite{c2,c8}.

Effective decisions combine data, models, and, perhaps most importantly, the experience of epidemiologists, infectious disease modelers, and public health officials (who we call experts in this work)~\cite{c9}. The importance of this collaborative expertise becomes clear when examining successful international responses to health emergencies. Following the 2010 earthquake in Haiti, the Centers for Disease Control and Prevention (CDC) supported the Haitian Health Ministry (MSPP) to strengthen disease surveillance systems and laboratory testing at health facilities already supported by the US President's Emergency Plan for AIDS Relief~\cite{c10}. When cholera was detected nine months later, the US Office of Foreign Disaster Assistance, MSPP, CDC, and local public health officials quickly coordinated the distribution of cholera treatment supplies to hospitals, promoted point-of-use water treatment and sanitation in displaced person camps and communities, and established a national cholera surveillance system still in use today~\cite{c11,c12}.

International responses like the above illustrate how expert judgment enables rapid decision-making under uncertainty---officials quickly assessed risk patterns, anticipated resource needs, and coordinated interventions. However, given the constraints imposed by reduced workforce capacity, maintaining the collaborative expertise demonstrated in past crisis responses will require innovative approaches to optimize what remains, and a deeper understanding of how expert judgment functions under pressure. To maximize the effectiveness of smaller teams, one must examine the complementary strengths of expert judgment and computational modeling, identifying how these approaches can be systematically combined to enhance forecasting accuracy and decision-making quality.

\begin{figure*}[h!]
\centering
\includegraphics[width=0.80\textwidth]{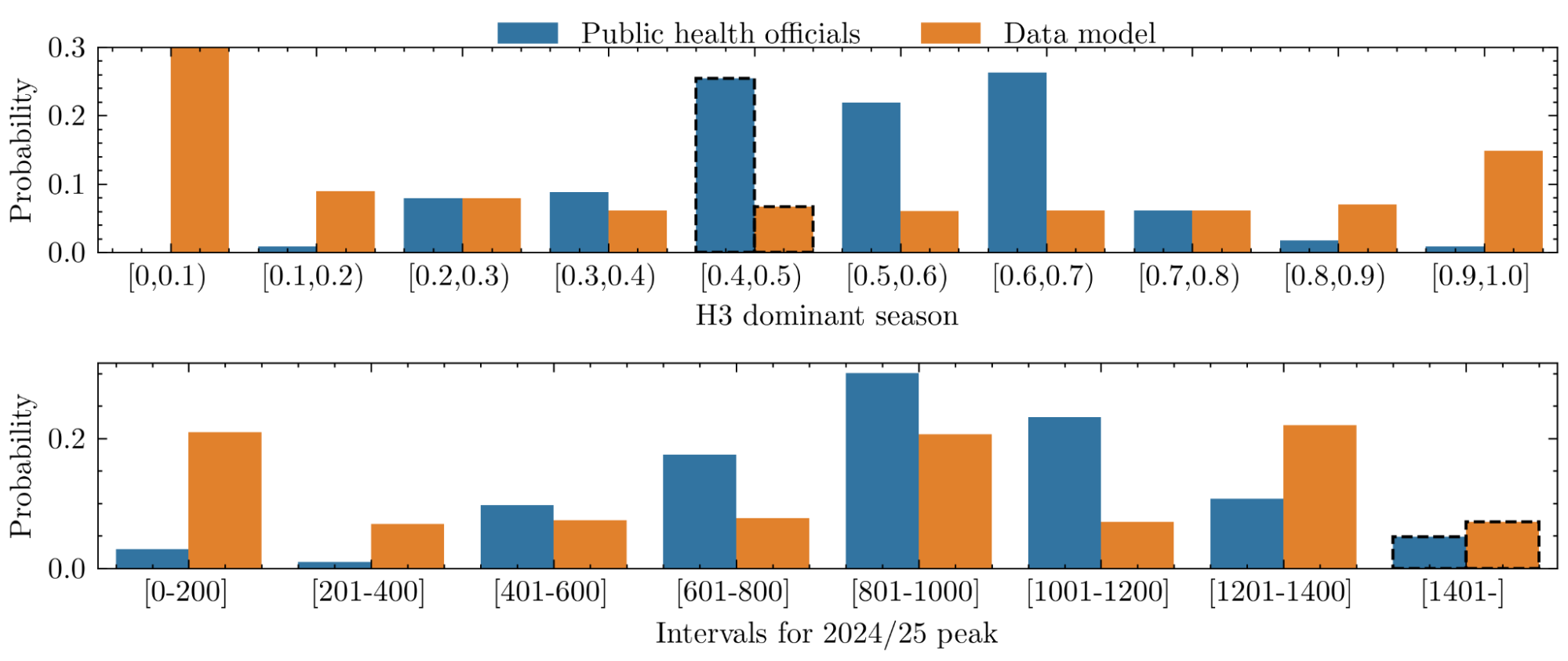}
\caption{Forecasts generated by a model (orange) and by a group of public health officials (blue) for two questions related to the 2024/25 influenza season in PA: (Top panel) would the majority of confirmed influenza cases be classified as H3 (vs H1) and (Bottom) what would be the peak number of incident hospitalizations. The height of the bars for both panels denote the probability assigned to each potential, future observation. The bars that are enclosed by dashed lines identify the ground truth which was computed at the close of the season. Public health officials generated forecasts of these two seasonal targets which were competitive with a traditional computational model.\label{figurelabel1}}
\end{figure*}

Past work has investigated the strengths and weaknesses of the ability of experts to produce well-performing predictions about the future~\cite{c13}. Past studies have shown that experts excel in assessing factors and mechanisms that are linked to worse health outcomes for a population~\cite{c14,c15}. In addition to assessing the link between factors and health conditions, when faced with a challenge (e.g. outbreak), experts typically make better decisions compared to a novice. This decision making ability is termed recognition-primed decision making. Given a challenge, the expert can quickly recognize past, similar experiences and select from a set of decisions that performed well~\cite{c16,c17}. That said, experts do not need to be the primary decision makers and can supplement statistical models. Experts have been asked to: design probability densities for models with sparse data, make predictions about specific aspects of an infectious outbreak (e.g. peak intensity, weekly cases, etc), and make long-range predictions about vaccine efficacy~\cite{c18,c19,c20}. However, expert judgment, like all human judgment, is susceptible to biases like group-think and anchoring~\cite{c21}. Expert predictions are not always as accurate as we would expect and may be overly confident~\cite{c22}. Importantly, expert judgment is crucial in the communication of information to the public.

In what follows, we contribute a case study comparing expert forecasts to two computational models, the ability of experts to generate reasonable mechanisms for their predictions, and recommend the necessary properties of a toolkit that could be used to collect expert predictions and rationales similar to how surveillance data is collected.

Our case study on expert forecasting of seasonal influenza examines how expert judgment functions in practice under a time constraint. On November 20th, 2024, at the council of State and Territorial Epidemiologists workshop on infectious disease forecasting, we posed two questions to public health officials, epidemiologists, and infectious disease modelers who attended the conference: in the state of Pennsylvania (PA), during the 2024/25 season, (1) will the majority of lab-confirmed cases of influenza be classified as H1 or H3 (H3 often results in more severe symptoms)? (2) what will be the peak number of confirmed hospitalizations due to influenza and why (e.g. a rationale for their prediction)? We received 217 responses from 114 experts (raw data provided in Supplement). Experts at the conference were provided brief background information to help them form a prediction and were given just five minutes to answer each question (Format for the questions can be found in the Supplement). The short timeframe to answer questions was meant to emphasize the role of public health experience over a more laborious study of background data. In addition to quantitative predictions, we analyzed expert rationales using Latent Dirichlet Allocation to identify themes in their reasoning.

\section{RESULTS}

\begin{figure*}[h!]
\centering
\includegraphics[width=0.80\textwidth]{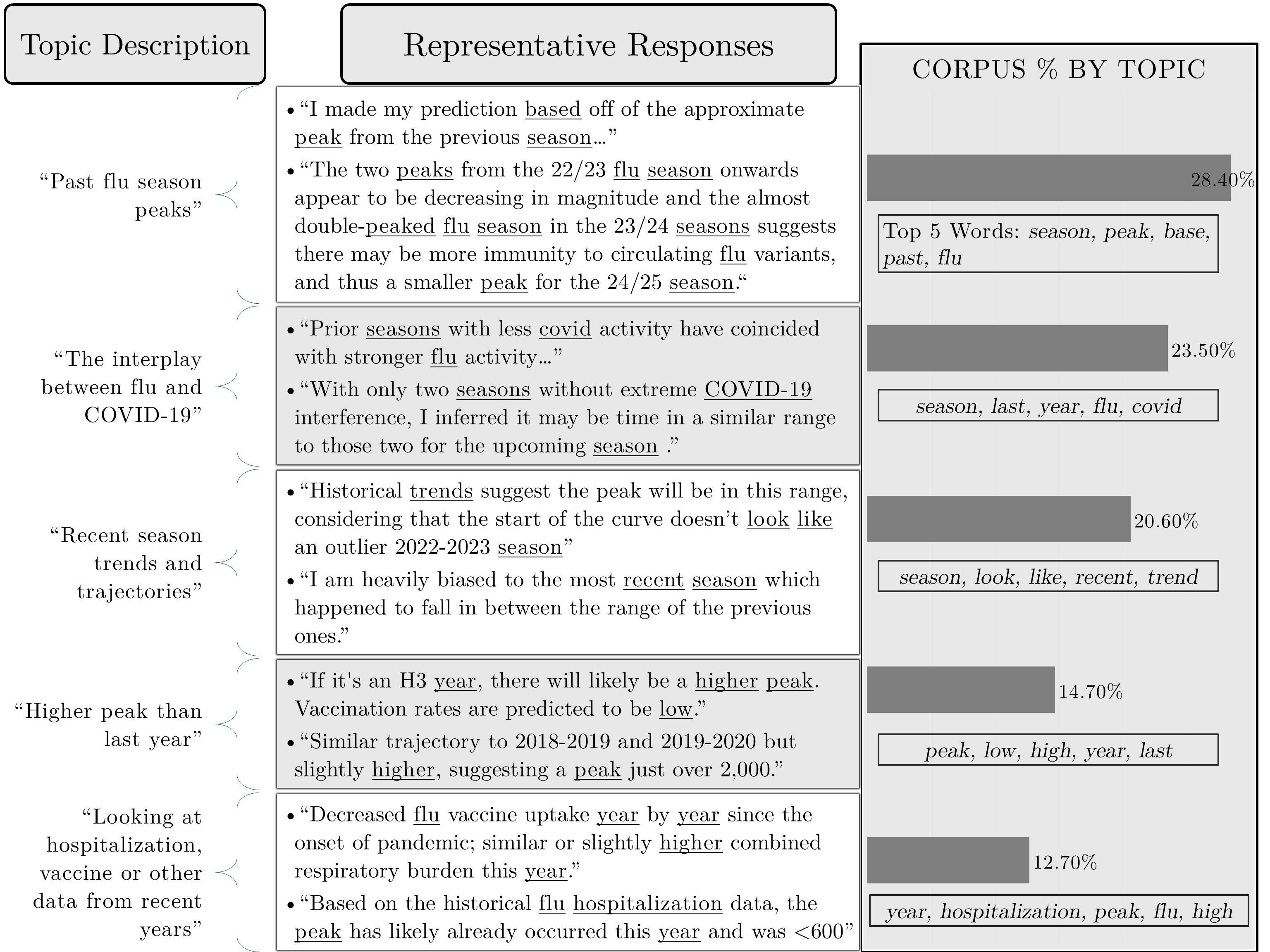}
\caption{Latent Dirichlet Allocation identified five topics from 102 expert responses, with prevalence shown as percentage of corpus. Topics demonstrate experts' consideration of historical flu patterns, COVID-19 interactions, recent trends, comparative assessments, and vaccination data---illustrating the diverse reasoning approaches that complement computational models in infectious disease forecasting.\label{figurelabel2}}
\end{figure*}

On May 31, 2025, at the close of the 2024/25 influenza season, the reported percent of H3 influenza cases in Pennsylvania was 40.0\% and in the US was 47.3\% (See Figure~\ref{figurelabel1}A). The percent of H3 influenza cases in Pennsylvania was on average 49\% over the past three seasons (84\% in season 2021/22, 53\% in 2022/23, and 20\% in 2023/24). Experts (vs the model) assigned a probability of 0.25 (vs 0.06) to the true percent of H3 influenza cases this season. In addition, experts (vs model) assigned a probability of 0.01 (vs 0.38) to an unlikely percent of H3 below 20\%. The model, though, did report a larger variance when compared to experts (model variance 11\% vs 2\% for experts).

In Pennsylvania, the peak number of incident hospitalizations for the 2024/25 season was 4,318 (See Figure~\ref{figurelabel1}B). This number of incident hospitalizations was the highest value compared to the past three seasons (200 in season 2021/22; 1,299 in 2022/23; and 933 in 2023/24). The model assigned a higher probability to the observed peak number of hospitalizations compared to the aggregate prediction of experts (model prob. = 0.07 vs experts prob = 0.05). That said, the model also assigned high probability to implausibly small peaks of less than 200 peak hospitalizations (model prob. = 0.21 vs Experts = 0.03).

When asked to provide a rationale for their prediction of peak incident hospitalizations, expert rationales focused on: historical flu patterns; how other pathogens like COVID-19 may modulate influenza intensity; scenarios such as if the season is (or is not) an H3-dominant season; and vaccine efficacy and uptake data. These topics were identified through Latent Dirichlet Allocation analysis of the rationales provided by experts (See Figure~\ref{figurelabel2}).

Several experts noted the interplay between COVID-19 and influenza, adding rationales like \emph{``with only two seasons without extreme COVID-19 interference, I inferred it may be time in a similar range to those two for the upcoming season''} and \emph{``Prior seasons with less covid activity have coincided with stronger flu activity. [...] I'd guess that we won't encounter a very high burden covid variant that displaces flu virus.''} Similar to the model, experts also considered previous peak data for influenza: \emph{``Looking at recent flu seasons, peaks have gone from about 1200 to about 1000. I am assuming there is a trend towards returning to lower peak incidence after reporting changes''} and \emph{``Prior flu seasons have peaks around 1K in January''}. Unlike the model, though, experts were able to tap a large breadth of knowledge on causal mechanisms that impact influenza: \emph{``Poor vaccine strain match will lead to a more severe influenza season than observed last year.''}, \emph{``Decreased flu vaccine uptake year by year since the onset of pandemic; similar or slightly higher combined respiratory burden this year.''}, and \emph{``last years average + bird flu''}. Some experts even tapped, in real-time, a math model, writing \emph{``Math model run in real time on the fly now''}. Notably, some experts were able to qualify their own potential biases when making predictions. One expert commented \emph{``When I saw the historical data from the link, I instinctively leaned [sic. towards this data] because my data scientist eye loves patterns, but : what if I'm wrong?''} as did another who wrote \emph{``I am heavily biased to the most recent season which happened to fall in between the range of the previous ones.''}

\section{DISCUSSION}

Our case study suggests complementary strengths between expert judgment and computational modeling in infectious disease forecasting. Public health experts' probability estimates for H3 influenza dominance aligned more closely with the observed outcome. While the forecast generated by the model and experts performed similarly for peak hospitalization predictions, the model assigned substantial probability to implausible scenarios that experts correctly identified as unlikely, such as peak hospitalizations below 200. A major advantage to collecting predictions from experts is their ability to describe their reasoning, while computational models only offer numerical output. These findings suggest that optimal forecasting approaches should collect both computational and expert forecasts (plus rationales), potentially leveraging natural language processing tools like small language models to systematically incorporate expert rationales into computational frameworks. Given just five minutes to produce a prediction and rationale, this work highlights the value of experience in public health practice.

This study has several important limitations. We lacked a direct comparison with non-expert predictions, limiting our ability to quantify the specific value of public health expertise versus general forecasting ability. Though the majority of attendees participated, not all conference attendees participated which could bias predictions toward individuals more confident or skilled in forecasting tasks. Our analysis focused on only two forecasting targets for a single influenza season in one geographic location which limited the scope for assessing expert performance across diverse scenarios or pathogens. Additionally, our simple computational model does not represent the full spectrum of sophisticated modeling approaches available, such as ensemble methods or machine learning techniques that might perform differently against expert judgment.

The strengths and susceptibilities of human judgment in forecasting aligns with established literature on expert decision-making under uncertainty~\cite{c13}. As demonstrated in our results and supported by prior research, experts excel at recognition-primed decision making, rapidly drawing on accumulated experience to assess plausible scenarios and causal mechanisms that purely data-driven models may miss~\cite{c16,c17}. However, expert forecasts remain vulnerable to cognitive biases, overconfidence, and anchoring effects that can compromise accuracy. This was exemplified in our work as smaller predictive variance for experts compared to a computational model. These biases may be particularly salient when experts are asked to make decisions in novel or rapidly evolving situations where past experience may be less applicable~\cite{c21,c22}.

Past forecasting systems that aggregate predictions and rationales from humans have found success that public health practice should explore. Modern forecasting increasingly relies on prediction platforms and specialized tools that leverage collective intelligence to enhance decision-making across diverse domains. Metaculus, an online forecasting platform, directly solicits probabilistic forecasts from its global community and aggregates these individual probabilities using sophisticated machine-learning-weighted predictions that outperform traditional prediction markets. Specialized tools like IDEAcology streamline rigorous expert elicitation through protocols designed for quantitative and probabilistic estimates in fields such as ecology and biosecurity~\cite{c23}. The success of existing forecasting platforms underscores the value of establishing formal systems for expert knowledge collection. For public health specifically, developing a dedicated web-based toolkit to coordinate national collection and analysis of expert rationales would treat human judgment as a critical data source alongside traditional surveillance systems. We posit a dedicated database of expert predictions, rationales, and the context in which they were made could improve crisis response.

To support public health response, especially during times of resource constraints, we recommend that future work address the construction of tools to amplify expert judgment, expertise, the decisions that were made and the results of those decisions. Expert reasoning should be considered a quantifiable data source on par with traditional data sources collected via surveillance systems.

\section{RECOMMENDATIONS}

We recommend the development of a toolkit that addresses the challenge of systematically capturing and leveraging expert judgment in public health forecasting. Although computational models provide quantitative predictions, they often lack the contextual reasoning and domain expertise that human forecasters bring to complex epidemiological scenarios. The proposed approach should combine expert forecasts with their underlying reasoning to create both improved predictive models and insights into decision-making processes.

Such a toolkit would require three components. First, the system should systematically collect forecasts from public health experts alongside structured documentation of their reasoning, assumptions, and the ground truth. Second, this system should employ a language model to identify characteristics of effective versus ineffective reasoning based on collected data from the first component. Third, a dashboard would present ensemble predictions, historical data, and a visual of common reasoning patterns across experts.

A toolkit such as the one we propose should be revised iteratively, with feedback collected from public health officials. This user-centered approach will ensure that data collection procedures, survey instruments, and dashboard interfaces align with existing workflows and decision-making needs in public health practice.

\section*{ACKNOWLEDGMENTS}

The authors thank Dr. Tomás Martín León (Modeling Section Chief, California Department of Public Health) and Justin Crow, MPA (Foresight \& Analytics Coordinator, Virginia Department of Health) for their valuable feedback.

\section{METHODS}

\subsection{Data Collection and determination of ground truth}

Participants were conference attendees at the Infectious Disease Forecasting Workshop hosted by the council of State and Territorial Epidemiologists (CSTE) and Centers for Disease Control and Prevention (CDC). The conference period was November 19-21, 2024. Conference attendees were experts in public health, epidemiology, and infectious disease modeling. Most, if not all, states were represented at the conference by active public health officials. We denote this group as experts.

On November 20, 2024 we posed two questions to experts during an invited talk. Question one asked ``Please assign a probability to this upcoming season being characterized as a H3 season in Pennsylvania''. Experts were able to provide the following answers: 10\%, 20\%,...,100\% probability that the season would be H3 dominant. Question two asked ``What will be the peak number of influenza hospitalizations in PA for the 2024/25 season?''. Experts were able to choose from a set of ranges: [0-200], [201-400], [401-600], [601-800], [801-1000], [1001-1200], [1201-1400], [1401-]. Brief background information was given to experts to aid them in forming predictions (see Supplemental).

Ground truth for both questions was determined on May 31, 2025, after the conclusion of the typical influenza season in the northern hemisphere (after MMWR week 2025W22). Ground truth on a H1 vs H3 dominant season was collected from the Pennsylvania Department of Health's Respiratory dashboard~\cite{c24}, and ground truth about the peak incident hospitalizations due to influenza was collected from the Weekly Hospital Respiratory Dataset which is hosted by the National Healthcare Safety Network as part of CDC~\cite{c25}.

\subsection{Evaluation}

Aggregate predictions from experts were compared to corresponding computational models that were trained on historical, observed data. The premise for this comparison is that if aggregate predictions from experts outperform reasonable computational models then these models, and the data that they are trained on, is not capturing information used by experts. This ``unknown'' information may include expertise, developed over years in the field.

\subsection{H1 vs H3 dominant influenza season}

Because the question asked to assign a probability to an H3 (vs H1) season, we can construct an aggregate density that assigns probability values to the future proportion of influenza cases that are typed as H3. Given $N$ predictions, the expert aggregate forecast assigned a probability to $x\%$ of cases typed as H3 equal to the number of experts who answered $x\%$ divided by $N$ which we denote $p_x$. We can build a probability density by assigning $p_x$ to the interval $[x,x+10\%)$. Our proposed computational model was a kernel density estimate that was trained on the past proportions of cases that were typed as H3 from the Pennsylvania Department of Health Respiratory Dashboard for seasons 2021/22 up until 2022/23 (See Supplemental for data set).

\subsection{Peak intensity of confirmed influenza hospitalizations}

The aggregate forecast from experts was a probability assignment to the above eight intervals outlined in the section titled \emph{Data Collection and determination of ground truth}. The probability assigned to interval $I$ was computed as the number of experts who selected that interval divided by all experts who participated. Our proposed computational model was a kernel density estimate that was trained on the past number of peak hospitalizations in Pennsylvania for seasons 2021/22 up until 2022/23 (See Supplemental for data set).

The rationales provided for predictions ($N$=102) were analyzed using Latent Dirichlet Allocation to identify common themes. Text preprocessing included removing stopwords, lemmatization, and filtering words appearing in <3 or >80\% of responses. After preprocessing, there were 61 unique words in any of the 102 rationales. Model selection involved training the LDA model for 2-8 topics and choosing the number of topics that resulted in the highest coherence scores. The highest coherence score (0.38) was when there were 5 topics selected. Final model used $\alpha$=0.083, $\beta$=0.01, trained for 100 iterations using Gensim 4.3.3. One author analyzed the topics (GH) and drafted interpretations independently, based on the words included and example responses. These interpretations were then reviewed and confirmed with a co-author (TM).

\section{Ethics Approval Statement}

In consultation with the Lehigh University Internal Review Board (IRB), this work was deemed not to involve human subjects and not need formal evaluation.

\section{Data availability and analysis reproducibility}

All data and code used to conduct the above analysis is available at https://github.com/computationalUncertaintyLab/cste\_predictions. In particular, a Makefile is provided that formats the data and then runs the analysis for this work.

\section{Supplemental Materials}

1. Line list of predictions and rationals from experts

2. Forms used for data collection (i.e. expert elicitation)

\addtolength{\textheight}{-12cm}

%%%%%%%%%%%%%%%%%%%%%%%%%%%%%%%%%%%%%%%%%%%%%%%%%%%%%%%%%%%%%%%%%%%%%%%%%%%%%%%%

\end{document}